# How Usable are Rust Cryptography APIs?

Kai Mindermann
University of Stuttgart
Institute of Software Technology
kai.mindermann@informatik.uni-stuttgart.de

Philipp Keck
University of Stuttgart
Institute of Software Technology

Stefan Wagner
University of Stuttgart
Institute of Software Technology
stefan.wagner@informatik.uni-stuttgart.de

*Abstract*—*Context:* Poor usability of cryptographic APIs is a severe source of vulnerabilities. *Aim:* We wanted to find out what kind of cryptographic libraries are present in Rust and how usable they are. *Method:* We explored Rust's cryptographic libraries through a systematic search, conducted an exploratory study on the major libraries and a controlled experiment on two of these libraries with 28 student participants. *Results:* Only half of the major libraries explicitly focus on usability and misuse resistance, which is reflected in their current APIs. We found that participants were more successful using *rust-crypto* which we considered less usable than *ring* before the experiment. *Conclusion:* We discuss API design insights and make recommendations for the design of crypto libraries in Rust regarding the detail and structure of the documentation, higher-level APIs as wrappers for the existing low-level libraries, and selected, good-quality example code to improve the emerging cryptographic libraries of Rust.

## I. Introduction

Rust is a systems programming language sponsored by Mozilla and was voted the most loved programming language in the 2016 StackOverflow developer survey [1]. It rivals other systems programming languages like C or C++ and despite being only seven years old with its first stable release in May 2015, Rust has already been used by Mozilla to build parts of the Firefox browser [2]. Apart from speed and concurrency, one of its main goals is safety, mostly achieved through an innovative memory management concept that makes it suitable for developing security-critical applications. In particular cryptographic libraries and services are thus likely to become widely used in Rust.

Many applications that use encryption or other cryptographic services end up being vulnerable because their developers knowingly or unknowingly misuse cryptographic libraries [3, 4, 5, 6]. While the cryptographic algorithms themselves may not contain severe bugs, their application programming interfaces (APIs) are not resistant to misuse and in some cases they are very difficult to use at all without detailed cryptographic knowledge. Such detailed knowledge cannot be expected from application programmers, as it is an integral part of software engineering to break down complex problems into smaller problems such that the person solving one of them does not need detailed knowledge of the other solutions. Therefore, many researchers have called for more usable and misuse-resistant crypto APIs [6, 7, 8, 9] and many make concrete recommendations for usable API designs [10, 11, 12].

The established crypto libraries such as OpenSSL cannot make significant changes to their APIs for compatibility reasons. Consequently, new libraries with a strong focus on API usability have been developed. Notable examples include NaCl, its offspring libsodium, Keyczar and the Python library *cryptography.io*.

Several major crypto libraries already exist in Rust including wrappers for OpenSSL (e.g. *rust-openssl*) and *libsodium*, some plain-Rust implementations like *rust-crypto* and a BoringSSL-based library called *ring*.

We see that cryptographic libraries often have usability problems. This work aims for analyzing how usable and misuse-resistant Rusts major crypto libraries are and how they have to be improved. We ask the following research questions:

**RQ 1:** *Which are the major cryptographic libraries in Rust?*

**RQ 2:** *How usable are the current Rust cryptographic libraries?*

To answer these research questions, we first conducted a systematic search for Rust crypto libraries. Next we performed an exploratory study on the major crypto libraries and their usability. Following that, we conducted a controlled experiment with two of the major libraries. From these we gathered insights and derived recommendations for improving the usability which we present at the end.

We contribute the following:
- Empirical insights into beginner usage and usability problems of Rust's cryptographic libraries.
- Recommendations for the design of cryptographic libraries in Rust to help improve their usability and misuse resistance.

## II. Background and Related work

To the best of our knowledge, the Research question RQ 1 (Which are the major cryptographic libraries in Rust?) has not been answered scientifically. In comparison, usability is a common research area. Therefore, we describe related work for the usability of cryptography APIs.

There are large bodies of literature regarding API design, API usability and crypto usability (e.g. the usability of encryption facilities by end-users). The usability of crypto APIs lies at their intersection, as it is fundamentally about making APIs easy to use and misuse-resistant but it also involves the same difficulty that cryptography is especially hard to understand and API users should not be faced with security-critical decisions unless it is unavoidable and they have been made aware of the consequences.

Georgiev et al. [3] analyzed SSL certificate validation code in a range of applications and found that the libraries themselves are correct "for the most part" but developers often misunderstand the APIs which is the "primary cause" for vulnerabilities. Egele et al. [5] automatically checked Android apps and found widespread security flaws such as ECB mode, constant keys, constant salts and constant pseudorandom number (PRNG) seeds, all of which can be traced back to a misuse of the cryptographic APIs in Android. Like many others, they proposed to get rid of insecure defaults (the ECB mode in Java's `Cipher.getInstance()`, for example) to increase misuse resistance, they recommended improving the API documentation and they proposed to build APIs which enforce their semantic contracts by means of the API itself. Lazar et al. [6] investigated 269 vulnerabilities from the CVE database and found that the majority is caused by application code which misuses the properly implemented cryptographic libraries. Das and King [7] defined seven properties to determine how safe a cryptographic library is and applied them to six libraries for the most popular programming languages. Nadi et al. [8] empirically investigated the Java Cryptography Architecture (JCA) by analyzing StackOverflow posts, GitHub repositories and surveying developers. They found that the APIs are perceived as being too low-level and recommended task-based API (similar to the *cryptography.io* library) and improved documentation as solutions. Acar et al. [13] compared the usability of cryptographic Python libraries in a controlled experiment with several tasks in an online environment. They found that simplicity is not enough, the documentation must be adequate as well. [10] developed 10 "principles for creating usable and secure crypto APIs" which are more or less directly applicable to Rust crypto libraries.

On a more technical level, Forler, Lucks, and Wenzel [12] demonstrated how the Ada language and compiler can be used to design an API that prevents nonce reuse and plaintext leaks. Similar constructions could be possible in Rust. There is no academic work regarding the usability of *Rust* crypto APIs but Devlin [14] gave a related talk at a community Meetup which was recorded. He mentions Too Many Choices, Unsafe Defaults, Unauthenticated Encryption, IV / Nonce Selection, In-SecureRandom, API Gaps and Password Hash as problematic topics. We want to address this research gap with findings from our exploratory study and a controlled experiment, both on the usability of the cryptographic libraries.

While Rust is a programming language gaining importance and being particularly suitable for crypto libraries, to the best of our knowledge, there has been no investigation of Rust's crypto libraries.

III. EXISTING LIBRARIES

Before we analyzed the usability of Rust's crypto libraries, we had to identify them and determined which are the most important ones.

To answer RQ 1 (Which are the major cryptographic libraries in Rust?), we used a systematic search approach to explore the population of Rust libraries, starting from the following places:
- Reddit thread "What crypto library do y'all use?"[1],
- the *crypto* and *cryptography* keywords on crates.io,
- the "Cryptography" collection on libs.rs[2],
- the *crypto* and *cryptography* topics on GitHub,
- an overview on `arewewebyet.org`[3],
- asking on the #rust-crypto IRC channel on the Mozilla IRC network.

Retracted libraries (with "yanked" crates[4]) were excluded. For every identified library, we also looked at other GitHub repositories of the main developers and found a few more libraries. As we intended to investigate and compare their usability, we grouped libraries that offer similar functionality. We first grouped them by the level of the provided algorithms and then subdivided the primitive level to split off smaller libraries with specific purposes from the larger, general-purpose libraries which we analyze in the remainder of this paper. Through our search, we found the following 81 libraries:
- 10 crypto-specific *utility* libraries for constant-time operations, secret memory and similar.
- 13 *larger* libraries which offer *multiple primitives* and usually multiple algorithms per primitive. The implementations are either written in Rust or attached through wrappers to code in another language. The major libraries in this category are introduced in section III-A.
- 35 libraries which implement a *single* primitive, algorithm or a small family of them in Rust.
- 5 libraries which offer a *simpler interface* to other implementations for specific application scenarios.
- 18 libraries that implement *cryptosystems or protocols* (mostly Transport Layer Security (TLS)) and usually depend on lower-level libraries.

We determined *major* libraries semi-manually: Any library with 20 or more dependent crates is considered a major library. We also added *octavo* because it implements multiple cryptographic primitives directly in Rust instead of wrapping another implementation and we included *rust_sodium* and *RustCrypto* – two forks of *sodiumoxide* and *rust-crypto*, respectively – in some of our analyses. All major libraries fall in the "multiple primitives" category.

*A. Major libraries for primitives*

Please see table I for an overview of the major Rust crypto libraries in chronological order. By far the oldest library is *rust-openssl* and like the wrapped C library OpenSSL, it offers both primitives and a TLS implementation. Its age and its double role explain the overall high number of commits, downloads and dependent crates. It is also the only library for primitives developed by a member of the official Rust library team (@sfackler). There has been a major overhaul in 2016 with about as many contributions as in all previous years combined.

---

[1] https://reddit.com/4d8hxm
[2] http://libs.rs/cryptography/
[3] http://www.arewewebyet.org/topics/crypto/
[4] *Crates* are Rust's compilation units (like packages).

TABLE I
KEY DATA OF MAJOR RUST CRYPTO LIBRARIES (AS OF 2 JUNE 2018)

| Name | rust-openssl | rust-crypto | sodiumoxide | octavo | ring | rust_sodium | RustCrypto |
|---|---|---|---|---|---|---|---|
| Based on | OpenSSL | (plain Rust) | libsodium | (plain Rust) | BoringSSL | sodiumoxide | rust-crypto |
| Started | Dec 2011 | Sep 2013 | Dec 2013 | Jul 2015 | Sep 2015 | Aug 2016 | Oct 2016 |
| Last commit | ongoing | Sep 2016 | ongoing | Oct 2017 | ongoing | Jun 2018 | ongoing |
| Last release | ongoing | May 2016 | ongoing | Apr 2016 | ongoing | May 2018 | ongoing |
| Commits | 2,325 | 753 | 534 | 290 | 8,490 (4,762) | 169 | 93 |
| Contributors | 167 | 49 | 41 | 11 | 128 (78) | 22 | 5 |
| Downloads | 1,763,450 | 662,891 | 95,258 | 1,053 | 323,635 | 23,049 | <20,000 |
| Downloads (60d) | 268,753 | 49,450 | 11,822 | 40 | 68,404 | 2,106 | (unclear) |
| GitHub stars | 372 | 725 | 313 | 123 | 865 | 45 | 21 |
| Dependent crates | 150 | 178 | 47 | 0 | 95 | 14 | 9+ |
| API level | medium | low | high | medium-low | medium-high | high | low |

Two projects aim to implement the most relevant cryptographic primitives natively in Rust, to eventually become an alternative to the established crypto libraries in C. The older project is *rust-crypto* which is the second most popular crypto library after *rust-openssl* and the newer one is *octavo* which is still incomplete and insecure. Both libraries have APIs that grew directly on top of the algorithms instead of another library's interface, resulting in relatively low-level APIs. Development on both has stalled since 2016, though there is a recent effort to revive and modularize *rust-crypto* under the GitHub organization *RustCrypto*. Because of this modularization, the statistics of *RustCrypto* could not be determined exactly (see table I).

*sodiumoxide* is a more usable wrapper API around the Rust bindings for *libsodium*. The functions and primitives offered by *libsodium* are rather high-level and originate from Daniel Bernstein's NaCl library. Like the underlying libraries, *sodiumoxide* is designed for usability and misuse resistance. *rust_sodium* is a fork of *sodiumoxide* that improves the build system and a few other things, making it easier to use especially on Windows, though it is not as actively maintained.

A relatively new library that aspires to provide even more usable and misuse-resistant APIs is *ring*. It uses a slimmed-down code base of BoringSSL, Google's simplified OpenSSL fork and excludes the entire TLS stack and most deprecated algorithms. The API is developed independently of the underlying Assembler/C code and notable efforts and innovations are made regarding misuse resistance. Because *ring* is technically a fork of BoringSSL, all contributions to BoringSSL are also counted towards *ring*'s statistics. The numbers in parentheses in table I quantify the contributions to *ring* that are not present in the upstream project. Even after these subtractions, *ring* still has by far the highest number of commits and quite many contributors considering its young age.

*B. Discussion*

Rust has surprisingly many and manifold cryptographic libraries. It seems that several programmers implement a small (crypto) project to explore and learn the Rust language. To answer RQ 1: Our systematic search identified seven major libraries that provide a range of relevant cryptographic primitives. The next question now is how usable these libraries are and whether we can offer suggestions for improvement.

IV. EXPLORATORY STUDY

In this section we describe an exploratory study with the major libraries to provide the first answers to RQ 2.

The second author of this paper (hereinafter called *explorer*) conducted an exploratory study to test several usability aspects of the existing crypto libraries: documentation, structure of the API, function signatures and misuse resistance. As a Rust beginner with significant programming experience in other languages and basic cryptographic knowledge, he completed a Rust tutorial and then started the exploratory study. It was conducted using the latest versions of the major libraries introduced in the previous section (see table I) in November 2016 and some of the uncovered issues have been fixed in the meantime. Note that *RustCrypto* did not exist at the time and *sodiumoxide* had the same API as *rust_sodium* but failed to build on Windows, so only *rust_sodium* was used. The explorer used *cargo* and *rustc* version 1.12.1 and IntelliJ IDEA with the *intellij-rust* plugin which provides code completion and navigation.

The exploratory study consisted of four stages designed to cover the most important APIs (hashing, HMAC, symmetric encryption, authenticated encryption (AE) and key management).

*A. Task*

A code skeleton was implemented before the actual exploration. That application spawns two threads, one (the *server*) opens a TCP port and the other (the *client*) connects. Then the client sends a string, the server confirms with a simple "1" and shuts down, the client waits for the confirmation to arrive and also shuts down causing the main thread to exit.

In the first stage, the respective crypto library was used to
(i) generate a symmetric encryption key and load it into both threads,

(ii) encrypt (and decrypt) the string sent across the TCP connection and
(iii) replace the simple confirmation with a hash value that has to match.

In the second stage, the encryption was replaced with authenticated encryption (AE). If the implementation from the first stage already had used AE or if no AE ciphers were available in the library, a different algorithm of the same type was chosen to test their substitutability.

In the third stage, the hash function was replaced with an HMAC using the same key as for the encryption. This is not recommendable from a cryptography point of view but it does not affect this exploration about API usability.

In the fourth stage, the protocol was kept unchanged but the received authenticators (the AE cipher's tag and the HMAC digest) were tampered with, simply by incrementing one of their bytes at the receiving end. The application's reaction to such a simulated attack was observed and a few alternative ways to call the API were investigated to find out how easily an invalid authenticator could remain unnoticed.

Depending on the encryption cipher, a nonce and/or an authentication tag needed to be sent to the server separately. The corresponding code was only developed once and copied for later explorations. Regarding the comparison of authenticators, it was evaluated how the libraries promoted or enforced the use of constant-time comparison functions.[5]

*B. rust-crypto*

Version 0.2.36 was used for the exploratory study.

*Documentation*: The documentation is partly detailed (e.g. for hashing), partly sparse (e.g. for HMAC) and partly nonexistent. Unfortunately, the symmetric encryption APIs are not documented at all but there is a code example[6] for AES-CBC in the repository (the only one).

*Signatures*: The API for AES-CBC requires the caller to feed the input and output through custom buffer types and react dynamically to buffer overflows or underflows. Considering the task at hand (encrypt a string entirely), the API is extremely low-level and difficult to understand, so that the explorer just copied the code example and used the higher-level API provided there. The AE, hashing and HMAC functions were relatively straightforward to use.

*API structure*: Block, stream and authenticated ciphers use entirely different APIs, as do hashing and HMAC. They are not independent, though, as *rust-crypto*'s API design cleverly integrates elements in a way that makes sense from an *implementation* perspective: the Hmac<D> takes a hashing algorithm (digest) D as a parameter, for example. The same concept is applied throughout the library, e.g., the generic CTR mode implementation can be used with any block cipher.

---

[5] At the time of the exploration, the explorer was not aware that the regular comparison functions (==) are insecure for these purposes—like most developers with only basic cryptographic knowledge. This evaluation was therefore done several months later.

[6] https://github.com/DaGenix/rust-crypto/blob/b6e3294/examples/symmetriccipher.rs

Hence, the trait-based API is not designed for ease of use or misuse resistance but to appropriately reflect the underlying algorithms and to make them composable and interoperable.

*Misuse resistance*: The HMAC API returns a custom struct which overrides the == operator so that comparisons are performed in constant time. When the authentication tag is tampered with, the decrypt() function returns false. The return value can be ignored without any compiler warnings but no harm is done because the function does not yield the encrypted plaintext unless the verification was successful.

*C. ring*

Version 0.5.3 was used for the exploratory study.

*Documentation*: The documentation is brief but complete. Most parts of the library have a code example, though the difficult AE API does not (yet).

*Signatures*: The encryption API requires allocating extra space for the authentication tag in the vector that is used for both input and output and finally truncating the result to the length returned from the encryption function. This API slowed the explorer down but did not result in any bugs or other problems. On the plus side, it saves the caller the trouble of dealing with a separate authentication tag. The hashing and HMAC APIs were easy to use.

*API structure*: There is one top-level module per primitive and all modules are structured similarly. In particular, they all provide top-level global functions for the main operations. Thus substituting the AES-GCM algorithm with a ChaCha20-Poly1305 was straightforward.

*Misuse resistance*: *ring* exclusively offers authenticated kinds of encryption which prevent accidental misuse of unauthenticated encryption. The HMAC module provides a verify() function that uses a constant-time comparison. Though its use is not enforced and there are no warnings on the getter for the raw digest data, the symmetrical design of the module and the code example strongly encourage using it. Tampering with the tag or digest produces errors in the return values of the decryption or verification functions and ignoring these return values leads to compiler warnings.

*D. rust-openssl*

Version 0.9.1 was used for the exploratory study. The exploration had been started with 0.9.0 but initial attempts to use the AES-GCM cipher failed because the wrapper API was not designed for AE ciphers. The library owner added the missing code within a few hours.[7]

*Documentation*: The library is partly documented but the documentation neither guides nor educates the user. For example, it does not explain which ciphers need a nonce or how long the nonces/keys need to be. The explorer struggled to find the HMAC implementation because the top-level documentation is mostly empty, there is no module named hmac and a documentation search leads to the pkey::PKey::hmac() constructor. Though this is the right struct to instantiate, there are no pointers towards the sign module that contains the necessary main functions. Once found, the sign module even has an extensive code example for HMAC which is one of many in the documentation.

---

[7] https://github.com/sfackler/rust-openssl/pull/519 Note how the introduction of this new API is immediately followed (though not preceded) by discussions about API design and misuse resistance.

*Signatures*: Although the functions are mostly easy to use, it can be difficult to find the right one (cf. finding the sign module for HMAC). There are separate functions for AE and non-AE ciphers but they both accept instances of the same Cipher struct and an *optional* nonce. Therefore, the compiler cannot prohibit nonsensical calls like using ECB mode with a nonce, CBC mode without a nonce, a non-AE cipher with the AE API or vice-versa. The former goes unnoticed entirely, a missing nonce leads to a good error message during encryption and the latter two just fail at encryption or decryption with a confusing message.

*API structure*: *rust-openssl* contains cryptographic primitives and a TLS implementation, though they are not noticeably separated. Some modules also use global functions for the main operations, whereas others require the user to instantiate a struct and call its functions.

*Misuse resistance*: Even though the signing API has a Verifier that could take care of the constant-time comparison, it does not work for HMACs. Not knowing about timing attacks, the explorer implemented the comparison with ==. When the AE tag is tampered with, the decryption function returns an error instead of the plaintext – the API uses a Rust enum to return either the plaintext or an error – which prevents any accidental misuse.

### E. rust_sodium

Version 0.1.2 was used for the exploratory study.

*Documentation*: The top-level documentation points straight to the relevant modules and every module contains relevant code examples.

*Signatures*: *rust_sodium* was by far the easiest to use because the functions are rather high-level, use strong types for their arguments (e.g. Key and Nonce) and there are generator functions for these types in the same module. This type safety also exposed the (actually bad) use of a single key for encryption and HMAC, forcing the explorer to create two instances from the same key material.

*API structure*: Every algorithm lives in its own module and the modules for the same primitive have exactly the same structure and are grouped in a top-level module which makes switching algorithms easy. Switching to another encryption cipher was not possible, however, because XSalsa20-Poly1305 is the only one. The many modules are generated by macros (to avoid code duplication) which confused the intellij−rust plugin so that code completion and navigation were not available in the exploratory study.

*Misuse resistance*: *rust_sodium* prevents many misuses through type safety and an opinionated selection of algorithms (inherited from *libsodium*) that excludes insecure algorithms and dangerous primitives like unauthenticated encryption. The documentation even advises against using the hash and auth (HMAC) modules "unless you know what you're doing." There is a verify () function with internal constant-time comparison that is advertised through its positioning and the code example, so that misuse is unlikely. When the AE tag is tampered with, the decryption function fails and thus prevents any accidental misuse.

### F. octavo

Version 0.1.1 was used for the exploratory study.

*Completeness*: *octavo* does not implement AES, so the ChaCha20 stream cipher was used in the first stage. The second stage had to be skipped for the lack of AE or any other stream cipher.

*Documentation*: Although the top-level documentation contains extensive information about how a cryptosystem is defined mathematically, about Kerckhoff's Principle and about what key lengths are secure, it does not inform about the key/initialization vector (IV) lengths actually required by the ChaCha20 cipher. Digging into the code (or trial and error) revealed that they must be multiples of 32 bits; hence the 112 bits recommended for "medium-term protection" by the documentation would not work. Otherwise, the encryption and HMAC documentations are basically empty, whereas the hashing part has good explanations, a code example and visible warnings about insecure algorithms.

*Signatures*: All APIs require preallocated vectors for output through &mut parameters but are otherwise easy to use.

*API structure*: *octavo* splits its functionality into several crates and re-exports them through a central crate. This can help reduce compile times and binary sizes but makes the code and documentation more difficult to navigate. Like *rust-crypto*, *octavo* uses generics to parameterize its Hmac implementation with any hash algorithm. This makes the components nicely composable but often requires the caller to use the unpleasant "turbofish" operator.[8]

*Misuse resistance*: *octavo* does not provide a random number generator (RNG) or a constant-time comparison function. The explorer used the *rand* crate instead and unknowingly implemented a == comparison vulnerable to timing attacks.

### G. General observations and summary

The explorer often had trouble preallocating the vectors for &mut [u8] parameters (Rust's out-parameters). There is a simple solution since March 2015: vec ![0; length ]. But before that many Rust users were frustrated when allocating a vector that some ridiculous solutions were suggested, including one that requires two lines of code and an unsafe block[9] (still the accepted answer) or one that creates an infinite iterator and collects it. The latter is still used in *rust-crypto*'s internal code today which is why the explorer ended up with that solution. In the case of *ring*, the input parameter is used for the output, too but it needs to be extended to make room for the authentication tag. Even *ring*'s own unit test[10] uses a loop to push the right

---

[8]https://twitter.com/steveklabnik/status/659034597062262784
[9]http://stackoverflow.com/a/28209155
[10]https://github.com/briansmith/ring/blob/master/src/aead/aead.rs#L359

number of zeros to the end, rather than using the resize () method.

As evidenced by the previous paragraph, the explorer often turned to the library's own source code and particularly its unit tests. It is reasonable to assume that other library users would do the same, especially when no code examples are provided. On the other hand, if suitable code examples were available, they would help the user enormously, as they serve many purposes at once that the documentation would otherwise have to fulfill individually: point to the right API to use, recommend sensible choices for algorithms and key lengths, explain the order in which the functions must be called and illustrate how the returned result can be used.

It should be pointed out that this explorative study is *not* representative of all Rust crypto libraries, as it only covers major primitive libraries. In particular, we did not investigate any TLS libraries or the like. By selecting the major libraries for the exploratory study, there is a bias towards further developed and polished libraries, though this bias appropriately reflects the experience of the average crypto user in Rust.

Regarding RQ 2 we can see that there are libraries that make a great effort to provide good usability (e.g. *rust_sodium*) including the documentation, code examples and misuse resistance. Still, the libraries (e.g. *rust-crypto*) that are not only good wrappers around existing other libraries (like *OpenSSL* or *libsodium*) lack in usability.

## V. CONTROLLED EXPERIMENT

The usability of the crypto libraries should not only be analyzed subjectively by a single person as in the previously described exploratory study. Therefor we decided to further evaluate the usability of the libraries with a controlled experiment. In this experiment, we compared two of the major cryptographic libraries using two groups who worked on the same simple cryptographic task but each group had to use a different library to solve the task. This gives us additional answers to RQ2.

### A. Design

Initially, we thought about comparing all of Rust's major libraries but due to the limited number of possible participants we had to choose two libaries: We skipped the oldest *rust-openssl* library because its API intentionally mirrors the OpenSSL API and instead chose the *rust-crypto* (0.2.36) library which implements its primitives in Rust and is not based on any other libraries, for the first group. Like *rust-openssl*, *rust-crypto* does not focus on usable API design. Quite the opposite, it is the main developer's goal to "focus on creating high quality implementations [...] with idomatic [sic], maximally powerful interfaces without making (too many) concessions to ergonomics [sic]" (DaGenix 2015). For the second group, we chose a more recent library called *ring* (0.6.3) which uses the same algorithm implementations (in C and assembler) as BoringSSL but provides an entirely independent Rust API. *ring* is an interesting candidate because the main developer wants it to be high-level and as usable and "foolproof" as possible. During the experiment Rust (`rustc`) was at version 1.15.0 (10893a9a3 2017-01-19).

### B. Task and code skeleton

Due to time constraints (1 h 30 min including exit survey) and experience from the exploratory study, we decided to only offer one task: *Add symmetric encryption to an artificial application*. The artificial application consists only of a main method that gets *business texts* from a separate method (line 20) and should encrypt these texts (*TODO* comments in line 27–31) and then decrypt (*TODO* comments in line 41–45) them again. The code skeleton for *rust-crypto* is shown below (the one for *ring* is very similar except for the referenced crates and `use`-statement for the random number generator (line 12).

```rust
#[macro_use] extern crate log;
extern crate log4rs;
extern crate rustc_serialize;
extern crate bufstream;
extern crate rand;

extern crate crypto;

mod business;

fn main() {
    use rand::{Rng, os};

    // Initialize the logger.
    // You can produce logging outputs like this:
    // info!("Here is a number: {}", 42)
    log4rs::init_file("config/log4rs.yaml", Default::default
        ()).unwrap();

    // Retrieve the business text that should be encrypted.
    let plain_business_text = business::
        get_long_important_business_text(2);
    info!("Plain text: {}", &plain_business_text);

    // Create a copy of the plain text for later comparison
    //     (do not use this variable in your code).
    let copy_business_text = plain_business_text.clone();

    /**
    * ####################################
    * # TODO Set up your ENCRYPTION here #
    * ####################################
    */

    // Create Key, for example with:
    // let key: [u8; <length in Bytes>] = os::OsRng::new().
    //     unwrap().gen();
    // For AES-256 the keylength is 256 bits, so 32 bytes.

    let encrypted_text: &[u8] = &[0; 0]; // Instead of &[u8
        ], Vec<u8> or String are also fine.
    info!("Encrypted text: {:?}", &encrypted_text);

    /**
    * ####################################
    * # TODO Set up the DECRYPTION here #
    * ####################################
    */

    let decrypted_text: String = String::new();
    info!("Decrypted text: {}", &decrypted_text);

    // Output whether it worked:
    info!("Original and decrypted texts match: {}",
        copy_business_text == decrypted_text);
}
```

## C. Hypothesis

Our empirical hypothesis follows from the selection criteria described above: We expect *ring* to have a higher effectiveness, efficiency and satisfaction than *rust-crypto* for Rust beginners. We measure these three sub-factors of usability individually as described in the following analysis sections. We derive the following statistical hypotheses:

$H_{0.effectiveness}$: There is no difference in effectiveness in using *ring* and *rust-crypto*.

$H_{A.effectiveness}$: *ring* usage is more effective than *rust-crypto* usage.

$H_{0.efficiency}$: There is no difference in efficiency in using *ring* and *rust-crypto*.

$H_{A.efficiency}$: *ring* usage is more efficient than *rust-crypto* usage.

$H_{0.satisfaction}$: There is no difference in satisfaction in using *ring* and *rust-crypto*.

$H_{A.satisfaction}$: *ring* usage is more satisfactory than *rust-crypto* usage.

## D. Sample

The controlled experiment was conducted with students of the University of Osnabrück, who had just completed a semester-long lecture on the Rust programming language.

Before we describe the sample, we exclude some of the participant's results because participants 3, 4, 6, 14 and 23 tried to implement the Advanced Encryption Standard themselves but failed. We ignore their data as they might have performed differently if the task had been understood better. For the same reason we discount participant 15 who implemented a Caesar cipher (unsuccessful). We also disregard the data from the supervisor (participant 13) of the course. This leaves 11 participants in both groups.

The remaining 22 participants were between 19 and 26 years old (median 22.5) and most of them were male (seven were female). Their course of study is mostly *computer science* (16) but six were enrolled in *cognitive science*. All participants described their knowledge of cryptography as *No experience* or *moderate experience*. This is good, because we wanted to see how programmers that are not very knowledgeable in cryptography get along with cryptographic libraries. The rating for the programming experience had most responses with *I did several projects with other programming languages* (9) and the Rust experience was described by most participants as *I have not used Rust before the lecture ...* (12).

## E. Effectiveness

We divided the task into several subtasks that had to be done to complete the overall task. These subtasks were weighted differently depending on their difficulty which was determined based on the experience from the exploratory study. The final weighting is annotated in our task solution code in section V-I.

We denote the overall effectiveness with $\overline{E} \in [0, 1]$ and calculated it as the mean of all participants' effectiveness. The effectiveness $E \in [0, 1]$ of a participant was calculated as the weighted sum of the completed subtasks. A subtask can either be completed or not.

Calculating the effectiveness for each participant and summing it up to get a total effectiveness of the experiment sample we got $\overline{E}_{\text{rc}} = 0.66$ for *rust-crypto* and $\overline{E}_{\text{ring}} = 0.28$ for *ring*. The distributions for the effectiveness of the two libraries is depicted in the top left boxplot in Figure 1.

The descriptive statistics already suggest that we will not be able to reject $H_{0.effectiveness}$. We used the Student's t-test (because a Shapiro–Wilk test showed that the distribution is normal in both groups and the mean of the variance is equal according to an F-test). The result of the t-test is a p-value of 0.9966 which is much larger than our significance level of 0.05. Hence, we could not reject the null hypothesis that there is no difference in the effectiveness of using *rust-crypto* and *ring*.

## F. Efficiency

The resources programmers can spend are manifold. A common resource is time spent. We denote the *time-based efficiency* with $\overline{P} \in \mathbb{Q}$ and calculated it as the mean of all participants time-based efficiencies. The time-based efficiency $P \in \mathbb{Q}$ of a participant is calculated by the effectiveness $E$ divided by the time to completion or, in case of not completing the task, divided by the time until the time limit was reached.

According to the definition, we calculated the time-based efficiency for each participant and then calculated the mean to get the overall efficiency. This led to an efficiency of $\overline{P}_{\text{rc}} = 0.88$ tasks/hour for *rust-crypto* and $\overline{P}_{\text{ring}} = 0.30$ tasks/hour for *ring*. The distributions are shown in the boxplot in the top middle of Figure 1.

Measuring the subtasks' time is very inaccurate in itself, because developers work on different subtasks at the same time or it is not clear on which one they are currently working. Therefore, we cannot report on the timing of subtasks.

The descriptive statistics again suggest that we cannot reject this null hypothesis ($H_{0.efficiency}$) either. We used the Wilcoxon signed-rank test (because a Shapiro–Wilk test suggests the sample distribution of *rust-crypto*'s group is not normal). Again the large p-value 0.9953 is larger than our significance level. Hence, we could not reject the null hypothesis that there is no difference in the efficiency of using *rust-crypto* and *ring*.

## G. Satisfaction

Satisfaction can be measured by standardized questionnaires. We used the System Usability Scale (SUS) by Brooke [15] (similar to [13]) which is simple, quick and accurate [16] as it contains only a set of 10 questions. It can be analyzed with a simple formula and gives a SUS score in the range 0 to 100. We denote the overall satisfaction with $\overline{S} \in [0, 100]$ and calculated it as the mean of all participants' SUS scores. An average SUS score is 68 [16].

The mean SUS scores for *rust-crypto* are $\overline{S}_{\text{rc}} = 50.68$ and *ring*'s mean SUS score is $\overline{S}_{\text{ring}} = 33.64$. The distribution of the satisfaction of the two libraries is shown in the top right boxplot in Figure 1.

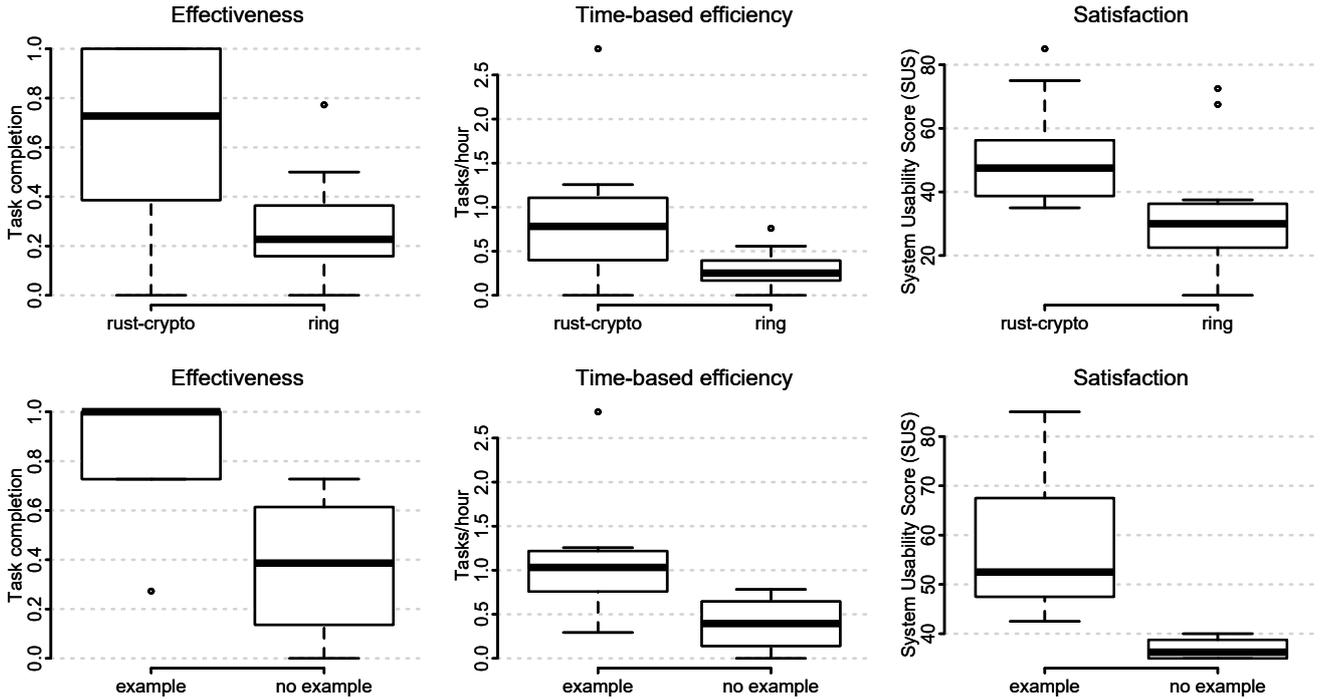

Fig. 1. Distributions of per-participant effectiveness, time-based efficiency and satisfaction, grouped by the used library (top row) and grouped by the participants who used the example and the ones that did not for *rust-crypto* (bottom row).

TABLE II
SUMMARY OF THE RESULTS FOR THE CONTROLLED EXPERIMENT

| Library | Effectiveness $\overline{E} \in [0,1]$ | Efficiency $\overline{P} \in \mathbb{Q}$ | Satisfaction $\overline{S} \in [0,100]$ |
|---|---|---|---|
| *ring* | 0.28 | 0.30 tasks/hour | 33.6 |
| *rust-crypto* | 0.66 | 0.88 tasks/hour | 50.7 |
| *rust-crypto* (with example) | 0.82 | 1.15 tasks/hour | 58.6 |
| *rust-crypto* (without example) | 0.38 | 0.39 tasks/hour | 36.9 |
| *PyCrypto* [13] | - | - | 63.9 |
| *Keyczar* [13] | - | - | 40.9 |

Also here, the descriptive statistics suggest that we cannot reject the third null hypothesis $H_{0.satisfaction}$. We used the Student's t-test (because a Shapiro–Wilk test showed that the distribution is normal in both groups and the mean of the variances is equal according to a F-test). The resulting p-value is 0.9792 and, hence, again far larger than our significance level. We could not reject the null hypothesis that there is no difference in the satisfaction in using *rust-crypto* and *ring*.

A summary of the results for the effectiveness, efficiency and satisfaction of *rust-crypto* and *ring* is shown in table II.

### H. Example code usage analysis

During the analysis we recognized that some participants used a suitable code example from the documentation/code repository of *rust-crypto*. Therefore, we compared the performance of the participants who used that example code with the ones that did not use it (just for *rust-crypto* participants). Only 4 of 7 (57%) who used the example code from/for *rust-crypto*, completed the task. Participants who did not use the example did not complete the task. We can calculate an effectiveness of 0.82 for participants who used the example and only 0.38 for the others. The efficiency is 1.15 tasks/hour with example code usage and 0.39 tasks/hour without using the example code. The mean SUS scores for *rust-crypto* are 58.57 with example code usage and 36.88 without using the example code. The distribution of the values is shown in the bottom row of Figure 1.

These observations let us conclude that it was mainly due to the example code that *rust-crypto* users performed better, as the participants who did not use the example code were equally unsuccessful completing the task as the participants who used *ring*.

### I. Task solution code and partial effectiveness

Here we show the correct solutions that participants had to program in the experiment. We also list the weights (percentage values in code comments) to get an idea of how we specified the difficulty for the mentioned subtasks that were used to create a more appropriate understanding of the effectiveness in subsection V-E. As explained, participants could use an example code for *rust-crypto*. Depending on the usage, there are two possible task solutions.

If the example code is not used, the task solution looks like this for *rust-crypto*:

```rust
let key: [u8; 256 / 8] = os::OsRng::new().unwrap().gen();
    // [keygen; 5%]
let iv: [u8; 128 / 8] = os::OsRng::new().unwrap().gen();
    // [nonce; 8%] (unless ECB is used)
// Instantiate encryptor with KeySize and padding [
    ecb_encryptor/cbc_encryptor; 12%]
let mut encryptor = aes::cbc_encryptor(aes::KeySize::
    KeySize256, &key, &iv, blockmodes::PkcsPadding);
// Allocate buffers [buffer; 20%]
let mut final_result = Vec::<u8>::new();
let mut read_buffer = buffer::RefReadBuffer::new(
    plain_business_text.as_ref());
let mut buffer = [0; 4096];
let mut write_buffer = buffer::RefWriteBuffer::new(&mut
    buffer);
// Call the block-wise API, possibly in a loop if the
    block size is fixed [blockcall; 40%]
loop {
  let result = encryptor.encrypt(&mut read_buffer, &mut
      write_buffer, true).unwrap();
  final_result.extend(write_buffer.take_read_buffer().
      take_remaining().iter().map(|&i| i));
  match result {
    buffer::BufferResult::BufferUnderflow => break,
    buffer::BufferResult::BufferOverflow => {}
  }
}

// Do the same for decryption (works exactly analogously)
    [decrypt; 10%]
let mut decryptor = aes::cbc_decryptor(aes::KeySize::
    KeySize256, &key, &iv, blockmodes::PkcsPadding);
...
let decrypted_text: String = String::from_utf8(
    final_result).unwrap(); // [from_utf8; 5%]
```

We can see for example that we assigned a weight of 40% for the loop block (lines 11–18) that repeatedly applies the needed encryption steps for a block-wise encryption method which must be used. As the concept of block-wise encryption must be understood and applied, we assumed (and observed in the screen recordings) a higher difficulty.

If the example code is used, the task solution looks like this for *rust-crypto*:

```rust
// Place a copy of the encrypt() and decrypt() functions
    from the code example into a separate module, into
// the main module or into the main() function (or
    anywhere else where they are accessible).

let key: [u8; 256 / 8] = os::OsRng::new().unwrap().gen();
    // [keygen; 20%]
let iv: [u8; 128 / 8] = os::OsRng::new().unwrap().gen();
    // [nonce; 30%]
let encrypted_text = encrypt(plain_business_text.as_bytes
    (), &key, &iv).unwrap(); // [call; 10%]

let decrypted_text = decrypt(&encrypted_text, &key, &iv).
    unwrap(); // [decrypt; 10%]
let decrypted_text: String = String::from_utf8(
    decrypted_text).unwrap(); // [from_utf8; 30%]
```

This solution requires visibly less code because it calls methods (`encrypt` and `decrypt`) which are copied from the example code and effectively offer a more convenient API for the participants who used the example code.

For *ring*, the task solution looks like this:

```rust
// [keygen, 5%]
let mut key_bytes = [0; 32];
SystemRandom::new().fill(&mut key_bytes).unwrap();
// nonce generation (participant needs to decide for
    random nonce) [nonce 8%]
let mut nonce = [0; 12];
SystemRandom::new().fill(&mut nonce).unwrap();
let sealing_key = aead::SealingKey::new(&aead::AES_256_GCM
    , &key_bytes).unwrap(); // [SealingKey, 7%]
// Allocate a buffer for the in_out parameter, filled with
     the plaintext and
// a further max_overhead_len empty bytes for the tag. [
    in_out, 40%]
let mut buffer = plain_business_text.into_bytes();
let newsize = buffer.len() + aead::AES_256_GCM.
    max_overhead_len();
buffer.resize(newsize, 0u8);
let encrypted_length = aead::seal_in_place(
&sealing_key, &nonce, &mut buffer,
aead::AES_256_GCM.max_overhead_len(), // set
    out_suffix_capacity to max_overhead_len [out-
    suffix, 5%]
&[] // set ad to an empty slice as there is no additional
    data [ad, 10%]
).unwrap();

let encrypted_text = &buffer[..encrypted_length]; //
    truncate the return value to the right length [
    truncate, 10%]

// Apply the knowledge gained from coding the above to
    implement the symmetrically structured decryption.
    [decrypt, 10%]
let opening_key = aead::OpeningKey::new(&aead::AES_256_GCM
    , &key_bytes).unwrap();
let mut buffer = encrypted_text.to_vec();
let text_length = aead::open_in_place(&opening_key, &nonce
    , 0, &mut buffer, &[]).unwrap();
// call from_utf8 to convert back to a String [from_utf8,
    5%]
let decrypted_text: String = String::from_utf8(buffer[..
    text_length].to_vec()).unwrap();
```

In this solution we assigned a high difficulty (weight 40%) to the allocation and initialization of a buffer of specific length that is later needed by *ring*'s `seal_in_place` method.

### J. General observations

Besides the quantitative data presented above, we made numerous observations when watching the screen recordings and also during discussions with participants after the experiment. Most of them only apply to a small number of participants because the overall group size was not large already but these issues appear common enough to probably affect other users, as well.

*Security*: Not a single participant was worried about the security implications of their choices during the experiment. Everyone went with the defaults provided by the library and tried to get something working. Those who succeeded did not reconsider earlier choices. Accordingly, the participants were rather unsure about the security of their code. In particular, all *rust-crypto* users ended up using unauthenticated encryption without knowing its potential dangers and the vast majority stuck to the CBC mode and PKCS padding given by the code example. The code example also made *rust-crypto* users more confident in the perceived security of their solution, whereas those who did not use it were as unsure as the *ring* users. This is in contrast to the fact that, if finished, the *ring* implementations would have been more secure because they forcibly use authenticated encryption. Besides one ECB usage, the only insecurity we found were nonces filled with zeros rather than random data (4 times).

*Nonce and other parameters*: *ring* users struggled to provide data for the nonce and ad parameters, as the documentation does not explain them. Besides finding the nonce length on the

Algorithm struct instances which most did, the participants had to figure out that it needs to be random or at least unique. 3 out of 8 participants who reached this point used zeros instead and one generated a new random nonce for the decryption. The ad parameter takes the additional data for AEAD ciphers which does not exist in this experiment and should be left empty (simply &[]). Some figured this out but others supplied seemingly arbitrary values to the parameter like the entire plaintext (again), the nonce or even the encryption key. Hence, a library's documentation should explain every parameter, especially if it is security-critical.

*Information sources*: Besides the official documentation of the libraries (briansmith.org, docs.rs and ironframeweork.io) and of Rust (rust-lang.org), a relatively large number of participants also visited Wikipedia and StackOverflow to learn about AES encryption, nonces and other topics. While some participants drove their web research through Rust-specific sites (mostly crates.io and docs.rs), others relied heavily on web search engines.

*Usage of code examples*: While a few participants copied the entire *rust-crypto* code example and then went ahead using the higher-level API that it exposes, the majority treated code examples differently. The participants tried to make sense of the code and apply it to their situation. Consequently, they initially only copied short sequences or typed analogous code themselves. This shows that code examples are treated differently from code hidden behind a library API: users want to understand it and they expect to adjust it. Therefore, code examples should be short and understandable. If a code example gets too long and even exposes its own API, as in the case of *rust-crypto*, it should better be part of the library's primary API, so that users do not worry about understanding it.

*In-place APIs are difficult*: Most *ring* users failed because of the time limit, that is, they did not head in the wrong direction but they had too many obstacles to overcome. The most time-consuming obstacles for *ring* users was its in_out parameter.[11] The library user needs to supply a byte slice to the encryption function which begins with the plaintext and has extra space at the end. Thus, participants had to convert the given string into a vector and resize it accordingly. The encryption function then overwrites the given data with the ciphertext and returns the length of the ciphertext. Thus, the participants had to use the returned length to truncate the vector and then treat it as the ciphertext. This kind of in-place API has a number of benefits: it does not require extra heap allocations on the part of the library – some environments do not have a heap, so this can be a hard requirement – and it uses minimal extra memory, as the plaintext can be overwritten. The particular format implemented by *ring* (appending the tag and supporting a "prefix" in the decryption function) is useful for certain applications. On the other hand, the usability of this API design is poor. The API exposed by the *code example* of *rust-crypto* is more high-level in comparison. Among other things, it works "out-of-place" and returns a newly allocated vector with the ciphertext. As only *rust-crypto* users who copied the code example were successful and *ring* users spent much time figuring out the in_out parameter, we conclude that such in-place APIs are unnecessarily complicated for the average user and should always be complemented with a higher-level, out-of-place API. At the time of writing, *ring* has already improved its existing API *and* the author is working on an out-of-place API, as well.

### K. Interpretation

Surprisingly, we found that the participants who used *rust-crypto* were more successful than the *ring* users. We could not reject a single null hypothesis. We assume this is largely due to the extensive code example provided by *rust-crypto* which is the only code example that the library has but it exactly fits the particular use case. Despite the random assignment of the library we found a difference between both groups regarding the Rust experience. The participants who used the supposedly less usable library *rust-crypto* had more experience with Rust than the other group. So this could also be the cause of the better performance.

The SUS scores for both *ring* (33.64) and *rust-crypto* (50.68) are well below the common average SUS score of 68 [16] and testify these libraries a bad satisfaction for the programmer. Slightly higher SUS scores were observed by Acar et al. [13] for python libraries of 63.9 for *PyCrypto* and 40.9 for *Keyczar* for example. Also, the effectiveness scores for the rather simple task of adding symmetric encryption to a program are low at 0.66 for *rust-crypto* and 0.28 for *ring*. Efficiency has similar differences. From the comments about the experiment we also get "Did I mention that I missed a good documentation for *rust-crypto*?" and "The *ring* library desperately needs a good documentation with examples". This suggests that the documentation has a lot of room for improvement.

### L. Threats to validity

Regarding the statistical conclusion validity, we are certain that we regarded the assumptions of the applied statistical tests and we did not violate the Type 1 and Type 2 errors in our hypothesis testing. Accordingly, we think our measures were reliable overall. As far as the arbitrary weighting and definition of subtasks go, we assume that others would define it in a similar way, not changing our results.

A big threat to our internal validity is the major factor for the unexpected performance difference: The available code example for *rust-crypto*. Despite its availability we think that this can also be seen as a major factor influencing the usability of the API as we discuss in subsection V-J. Another threat is the screen recording which was necessary for our analysis. We mitigated concerns by describing that the recordings will not contain any personal or other data that lets us connect it to a specific participant.

---

[11] Note that the documentation was not as detailed at the time. The version used by the participants is: https://docs.rs/ring/0.6.3/ring/aead/fn.open_in_place.html

The external validity is jeopardized at least for the generalization. The participants were students who had little programming experience. Regarding cryptographic knowledge we argue that this should be observable with more experienced programmers as well, so this is no big threat for the conclusions we draw. The time constraints limit the generalizability, because in reality programmers can spend more time on solving such a task.

## VI. DISCUSSION AND RECOMMENDATIONS

Viewing the current Rust crypto APIs in the light of recent research, we found that: Insecure defaults do not occur and most APIs try to avoid defaults entirely. Authenticated encryption is not advertised enough in low-level libraries, whereas the high-level libraries omit unauthenticated encryption altogether for maximal misuse resistance. Few high-level libraries are available. A few projects do not warn about deprecated/broken algorithms. There are no measures against accidental nonce reuse. Documentation and example code range from absent or scarce in the low-level libraries to excellent (in *sodiumoxide*).

These usability problems can be remedied in several both technical and non-technical ways. We recommend the following:

1) **Use a prominent location to link to the documentation.** E.g. at the start page of the repository.
2) **Create the first line of the module description carefully** because it is used in the summary of modules.
3) **Do not explain cryptographic concepts yourself** but link to comprehensible resources.
4) **Mention scenarios where the algorithm should be used** because non-experts might not select the appropriate algorithm for their use case.
5) **Mention closely related keywords**, as developers have different knowledge about cryptographic terms.
6) **Point out known weaknesses and vulnerabilities**. Even if the crypto API still offers weak or vulnerable algorithms, they should be clearly marked as such to prevent usage.
7) **Explain all parameters**. A lot of mistakes are made by providing wrong or insecure parameters. E. g. `key`: length in bits and bytes (the former is easier to recognize and remember, the latter is what users need in their code) or where the length can be found, possibly a pointer to a key generation function.
8) **Make recommendations when there are multiple choices**. E.g. if parameters can be constructed differently, it should be explained what the different choices imply.
9) **Provide high-level APIs for common scenarios**. Similar to the recommendation by Green and Smith [10] that non-crypto APIs should include and/or hide cryptographic functionality, a step in this direction is that the crypto APIs do this themselves.
10) **Promote the use of constant time comparison**. In Rust this can be done by implementing the PartialEq trait to override ==).
11) **Hide low-level APIs in a separate API layer called "hazardous materials"**. By naming it like this, developers take notice that they might be doing something dangerous.
12) **Provide example code for the most important use cases at module level**. As can be seen in our experiment the usability and comprehension of the API can be improved by offering example code that demonstrates the usage.
13) **Maintain example code the same way as security-relevant code**, as examples are a common entry point to understand the API or to quickly get to a solution. If they lack behind the API, the changes in the API might not be noticed or understood by developers.

Compared to Green and Smith's top ten principles [10], our recommendations are more specific but do not conflict with their suggestions.

## VII. CONCLUSION AND FUTURE WORK

The usability of cryptographic APIs has a significant impact on the security of the applications using them. As APIs are designed in the early stages of a library, it is the right time to investigate the young, emerging crypto libraries for the Rust programming language.

RQ 1 (Which are the major libraries in the ecosystem?) is sufficiently covered by our systematic search in section III. The major Rust libraries for cryptographic primitives are: *rust-openssl*, *rust-crypto*, *sodiumoxide*/*rust_sodium*, *octavo*, *ring* and *RustCrypto*. Regarding RQ 2 (How usable are the current Rust cryptographic libraries?) we learned from the exploratory study that the two oldest and most widespread libraries *rust-openssl* and *rust-crypto* have an API that is intended to be low-level and powerful sacrificing usability. But there are equally many libraries with a clear focus on usability. *sodiumoxide* and thus *rust_sodium* inherit the high-level interface from the underlying *libsodium* library and augment it with a "rustic" API that makes visible efforts towards usability. *ring* is based on the lower-level BoringSSL library but makes usability an explicit design goal and sacrifices backward compatibility for it. Currently, *ring* is not quite as easy to use as *sodiumoxide* but it is already relatively misuse-resistant and under heavy development to improve its usability. Consequently the usability and misuse resistance of the current libraries also varies, with the most usable libraries for cryptographic primitives being *sodiumoxide* and *ring*. Still, the controlled experiment found a rather bad usability for both the usability-focused *ring* and *rust-crypto*. Especially *ring* that had the explicit goal of usability performed worse than *rust-crypto*. The counterintuitive result of our experiment is that none of the *ring* users completed the task, whereas *rust-crypto* users were successful in 4 out of 11 cases. This success, however, was mostly only due to a fitting code example provided in *rust-crypto*'s repository and we found no relevant differences between *ring* users and *rust-crypto* users who did not use the code example. Nevertheless, having fitting code examples for common use cases is important. As often the examples are used to explore the API faster. Additionally, interesting observations were made about the developers. E. g. they are not worried about their choices regarding the security and many were slowed by obstacles originating from the crypto API. We derived major recommendations from these observations and the exploratory study which concern the detail and structure of the documentation, higher-level APIs and selected, good-quality example code.

The next step for us is the proposition of good example code for the major Rust cryptographic libraries. We intend to provide the examples directly within the documentation of the libraries but also via the CryptoExamples [17] platform. Further, we are in touch with the developers of the libraries to implement the recommendations that we made.

ACKNOWLEDGMENT

We would like to thank all reviewers and especially the active community members on the #rust-crypto IRC channel and @briansmith for your constructive feedback and good recommendations, insights and questions that all have made this work better. We would also like to thank Lukas Kalbertodt and Friedhelm Hofmeyer for providing us with the opportunity to conduct the experiment in parallel to their Rust lecture. Without them, we would not have been able to find so many rust programmers who could have participated in the experiment. This work was funded by the Baden-Württemberg Stiftung, thank you.